\documentclass[twocolumn,a4]{revtex4}
\usepackage{graphicx}
\usepackage{epstopdf}
\usepackage{longtable}

\begin{document}

\title{Solving the Einstein-Podolksy-Rosen puzzle: a possible origin of non-locality.}

\author{Werner A. Hofer}
\affiliation{Department of Physics, University of Liverpool\\
L69 3BX Liverpool, United Kingdom}

\begin{abstract}
So far no mechanism is known, which could connect the two measurements in a Bell-type experiment with a speed beyond the speed of light, commonly considered the ultimate limit of propagation of any field-like interaction. Here, we suggest such a mechanism, based on the phase of a photon field during its propagation. We show that two measurements, corresponding to two independent rotations of the fields, are connected, even if no signal passes from one point of measurement to the other. The non-local connection of a photon pair is the result of its origin at a common source, where the two fields acquire a well defined phase difference. Therefore, it is not actually a non-local effect in any conventional sense.
\end{abstract}\pacs{03.65.Ud}

\maketitle

One of the most puzzling results in modern physics, based originally on a paper by Einstein, Podolsky, and Rosen (EPR)\cite{EPR},
is the apparent non-locality of correlation measurements in quantum optics \cite{aspect99}. The measurements performed on pairs
of entangled photons, beginning with the experiments by Alain Aspect in 1982\cite{aspect82}, seem to prove beyond doubt that the
two measurements are not independent. The measurements are usually interpreted in terms of the Bell inequalities\cite{bell64},
which assert that their violation, corresponding to the experimental results and also the theoretical predictions of quantum
mechanics, amounts to a non-local connection between the two independent measurements[2].
\begin{figure}
\centering
\includegraphics[width=\columnwidth]{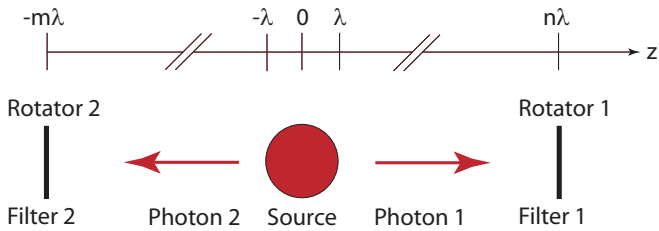}
\caption{(Color online) Bell-type experiment. Two photons are emitted from a common source, and subjected at two defined points $-m\lambda$ and $n\lambda$ to separate measurements, here assumed to be a rotation of the fields and a filtering process.}\label{Fig1}
\end{figure}
In Figure \ref{Fig1} we show the setup of a Bell-type experiment. Two photons with a defined phase difference $\Delta$ are emitted
from a common source. After traveling an integer number of wavelengths $\lambda$ of their associated field, either {\it n}, in
positive $z$-direction, or $m$, in negative $z$-direction, they are subjected to a measurement, which is assumed to consist
of a rotation of the photon fields. Rotations in geometric algebra\cite{lasenby02} are described by a multiplication with a
geometric product of two vectors. Here, we assume rotations perpendicular to the direction of photon propagation, which act
on a Poynting-like vector of the electromagnetic fields. The rotations
are then described by:
\begin{eqnarray}
R\left(z_1\right) &=& \exp \left({\bf e}_1{\bf e}_2\right) {\bf e}_3 z_1\,2 \pi/\lambda \nonumber \\
R\left(z_2\right) &=& \exp -\left({\bf e}_1{\bf e}_2\right) {\bf e}_3 z_2\,2 \pi/\lambda
\end{eqnarray}
where the values of $z_i$ are limited by $0 \le z_i \le \lambda$. The rotations thus cover all values from zero to a full
rotation of 2$\pi$. It is evident that the rotations are local measurements, i.e. the rotation at point $-m\lambda$ is independent
of the rotation at point $n\lambda$. Given that the geometric products involve a product of the three frame vectors ${\bf e}_i$,
the brackets can be omitted and the triple product $\left({\bf e}_1{\bf e}_2\right) {\bf e}_3 = {\bf e}_1{\bf e}_2 {\bf e}_3 = i$.
The two rotations are thus:
\begin{eqnarray}
R\left(\varphi_1\right) &=& \exp i z_1\,2 \pi/\lambda = e^{i\varphi_1} \nonumber \\
R\left(\varphi_2\right) &=& \exp -i \left(z_2\,2 \pi/\lambda + \Delta\right) = e^{-i\left(\varphi_2 + \Delta\right)}
\end{eqnarray}
where we symbolized the product $z_i 2\pi/\lambda$ by $\varphi_i$. The normalized probability $p$ of detecting photons after
a rotation with angle $\varphi_i$ shall be given by the square of the real part of the rotation, or:
\begin{equation}
p(\varphi_i) = \left(\Re (\varphi_i)\right)^2
\end{equation}
The probability in this case models a filter, acting after the rotator. Here, the measurement depends on the phase difference
between the source of the photon pair and the end point of the rotation. The real part of the phase difference is thus the square
root of the detection probability. For coincidence measurements we have therefore to consider the phase difference between the two
end points of the rotation of both photons. The correlations between two measurements with angles $\varphi_1,\varphi_2$ are then
described by a square of the real part of the product $R(\varphi_1) \cdot R(\varphi_2)$. The relations between rotations and
probabilities of photon measurements are:
\begin{eqnarray}
p(\varphi_1) &=& \cos^2 \varphi_1 \qquad p(\varphi_2) = \cos^2 \varphi_2 \nonumber \\
p(\varphi_1,\varphi_2) &=& \cos^2 (\varphi_1 - \varphi_2 - \Delta)
\end{eqnarray}
It is evident that in this case the correlations cannot be composed of a product of the individual probabilities.
The reason for this feature is that the rotations are complex phases, while the individual measurements are real values;
the product of probabilities thus does not account for the full connection of rotations in three dimensions. It is also
evident that the correlations describe the conditional probabilities in quantum mechanics, even though the individual
rotations are fully local. From this perspective it seems difficult to reconcile the framework with the local
derivations by John Bell\cite{bell64}, which, as well known, assume that locality implies such a separation of
local probability values, and that the product of these local probabilities is equal to the conditional probability
or the correlation probability. For rotations and complex phases such a derivation seems not justified.

The framework can be generalized to three and more rotations. Assuming that we have two rotators on either side,
positioned at integer values of the photon wavelength, the conditional probability for four individual measurements
with rotators  $\varphi_1$ to $\varphi_4$, where $\varphi_1$ and $\varphi_2$  are in positive $z$-direction while
$\varphi_3$ and $\varphi_4$ are in negative $z$-direction, is equal to:
\begin{equation}
p(\varphi_1,\varphi_2,\varphi_3,\varphi_4) = \cos^2 (\varphi_1 + \varphi_2 - \varphi_3 - \varphi_4 - \Delta)
\end{equation}
To appreciate the novelty of the approach it is illuminating to cite Alain Aspect's review paper in 1999
\cite{aspect99}: ''The violation of Bell's inequality, with strict relativistic separation between the chosen
measurements, means that it is impossible to maintain the image '\'a la Einstein' where correlations are explained
by common properties determined at the common source and subsequently carried along by each photon. We must
conclude that an entangled EPR photon pair is a non-separable object; that is, it is impossible to assign
individual local properties (local physical reality) to each photon. In some sense, both photons keep in
contact through space and time.''

Here, we found that the ''common property ... carried along by each photon'' is a complex phase, which
will be altered in a rotator. The actual normalized count $p$ does not reveal the full physical situation;
it is therefore necessary to take the correlated normalized count for the product of two complex rotations
and not, as assumed in the derivation of Bell's inequalities, the product of the two separate normalized
counts. The additional information about the imaginary component of the phase is not revealed in the local
counts, even though it is present in the local rotations. It seems thus that the difference between the
physical situation (a rotation of the fields), and the actual measurement result (a count of photons after
rotation), has not been appreciated to date. Rotations in three dimensional space, formalized within the
framework of geometric algebra, are the key to understanding spin properties of electrons, as recently
established\cite{hofer11}. Based on this analysis, it seems that they are equally key to understanding
polarizations of photons and electromagnetic fields.

Experimentally, the measurements are performed on a pair of down-converted photons\cite{aspect82}, which are
separated, subjected to a polarizer which can be interpreted as a combination of a rotator plus a filter,
and then measured either in a spin-up or spin-down state at the detectors. Experiments are usually interpreted
in terms of the Clauser-Horne-Shimony-Holt-inequalities (CHSH)\cite{CHSH}, which are based on normalized expectation
values $E(\varphi_1,\varphi_2)$, derived from coincidence counts of photon spins at the two points of measurement.
Within the present context it is actually unnecessary to define exactly, what spin-up and spin-down means in a
measurement; it suffices to assume that they will be subject to the same relation between rotational angles
and detection probability. For a phase-difference $\Delta = 0$ the normalized detection rates for spin-up and
spin-down photons will be (we denote coincidences by a capital C, as is standard in the literature, and also use
the convention that a coincidence is the measurement of equal spin for both, spin-up (+) and spin-down (-) components):
\begin{eqnarray}
C^{++} &=& C^{--} = \cos^2\left(\varphi_1 - \varphi_2\right) \nonumber \\
C^{+-} &=& C^{-+} = 1 - \cos^2\left(\varphi_1 - \varphi_2\right)
\end{eqnarray}
Then we obtain the same correlations of polarizations as in Aspect's first experiments\cite{aspect82}, namely:
\begin{equation}
E(\varphi_1,\varphi_2) = 2 \cos^2\left(\varphi_1 - \varphi_2\right) - 1 = \cos 2\left(\varphi_1 - \varphi_2\right)
\end{equation}
Correlations at different pairs of angles $\varphi_1,\varphi_2$ can be combined to a sum $S$, which, according to
Bell's derivation\cite{bell64}, should not be larger than two for any local model. Within the present model we
obtain, in accordance with experimental results and also with predictions of quantum mechanics:
\begin{eqnarray}
S\left(\varphi_1,\varphi_1',\varphi_2,\varphi_2'\right) &=& E(\varphi_1,\varphi_2) - E(\varphi_1,\varphi_2') + \\ &+&
E(\varphi_1',\varphi_2) + E(\varphi_1',\varphi_2') = 2 \sqrt{2} \nonumber
\end{eqnarray}
if  $\varphi_1 = 0, \varphi_1' = 45, \varphi_2  = 22.5, \varphi_2' = 67.5$, in violation of the Bell inequalities.
The model thus fully accounts for experimental values under ideal conditions (which are nearly reached in the most
advanced experiments\cite{weihs99}), and also for the standard predictions in quantum mechanics.

The underlying reason that quantum mechanics appears to be non-local is due to its formulation in terms of operators
and expectation values which entail integrations over the whole system. A local model, based on geometric algebra and
phases, can obtain the same numerical results, as shown here concerning EPR-type experiments.
Moreover, while the standard model makes the actual connection of entangled
photons somewhat less than transparent, the model developed has the advantage that all processes are local and transparent.
There is, as shown, no connection between the two measurements exceeding the speed of light. Moreover, the present
model is also a fully local model of photon entanglement. All that is required for entanglement, it seems, is a coherent
phase between the two photons. Whether this model is the whole answer to the problem, or only part of an eventually fully
comprehensive theory, cannot be estimated at present. This will to a large extent depend on subsequent experimental tests.
However, any claim that quantum entanglement is entirely incomprehensible and mysterious and forces us to renounce all
aspirations to modeling it in terms of standard physical concepts like realism and locality, an opinion frequently found
in the EPR literature, will seem hard to defend on the basis of this work.

From a statistical point of view an initial phase $\varphi_0$ at the origin for both photons does not alter the outcome.
Neither the single probabilities $p(\varphi_1)$ or $p(\varphi_2)$, nor the conditional probability $p(\varphi_1,\varphi_2)$
will be affected. In this context it is interesting to note that the present analysis is based only on the field properties of
photons and their rotational features. It does not need to consider any particle properties to arrive at the derived results.
However, it does also not specify, whether an individual photon at a particular setting of the polarizer will actually be
measured or not. For the macroscopic outcome such a specification is neither necessary, nor does it form part of any
theoretical model which describes the experiments at present. It is certainly not part of quantum mechanics, which, as
shown, can be replicated with a model based on three-dimensional rotations and phases of the photon fields. But it is
also not contained in Bell's analysis of the original EPR problem \cite{bell64}, where it is never specified, what will trigger the
detection of individual photons. Eventually, this question might be answered by a detailed analysis of the dynamical
processes in the polarizer (rotator/filter in the present model) itself, which contains hidden variables in the exact
shape of the fields or the thermal fluctuations of the polarizer atoms. Such a detailed picture is not necessary, though,
to establish the correlations which have been puzzling physicists for more than thirty years.

Returning to the original EPR problem and the question whether there is an ''element of reality'' in the experiments,
which is not described by the formalism in quantum mechanics, it turns out that Einstein was wrong: the description
in quantum mechanics via Pauli matrices is an equivalent way of accounting for rotations in three dimensional space,
as already emphasized \cite{hofer11}. It derives from the fact that the geometric algebra of three dimensional space
is equivalent to the algebra of the Pauli matrices:
\begin{eqnarray}
{\bf e}_i {\bf e}_j = \delta_{ij} + i \epsilon_{ijk} {\bf e}_k \nonumber \\
\hat{\sigma}_i \hat{\sigma} _j = \delta_{ij} + i \epsilon_{ijk} \hat{\sigma}_k
\end{eqnarray}
 The formalism in quantum mechanics is thus complete. However, Einstein was also right, because the
imaginary component of the phase difference, which is a consequence of rotations in geometric algebra, is not strictly
speaking a physical property of the system in quantum mechanics, and it is not revealed in the experiments, where only
the real component shows up in the photon count. It is thus a hidden variable. But this imaginary component is a - classical -
geometric component of rotations in geometric algebra, thus it does have physical reality. This reality has so far been
denied in quantum mechanics. In this sense one could say that even though the framework is formally complete, its relation
to physical reality has not been correctly described. This seems to account also for Bell's derivation of his famous
inequalities \cite{bell64}: since Bell did not ascribe physical reality to the imaginary component of the phase, he
did arrive only at a limited description of the situation from the viewpoint of geometric algebra. Thus his inequalities
can be violated in a local and realistic model, even though it has frequently been asserted that this is not possible.

On a final note we hope that this model and the clarifications presented in this paper will contribute to a more
rational debate about scientific issues in quantum mechanics in the future. Irrational statements about either the
scope or the epistemological meaning of quantum mechanics, which is after all only a scientific theory, seem
to have been the main obstacles to scientific progress in the past.

{\bf Acknowledgements} Helpful discussions with Konstantin Lukin and Sheldon Goldstein are gratefully acknowledged.
The author also acknowledges support from the Royal Society London.

\end{document}